\documentclass[11pt,english,a4]{article}
\usepackage{graphicx}
\usepackage{amstext}
\usepackage{caption}
\usepackage{etoolbox}
\usepackage{makeidx}
\include{epsf}
\usepackage{amsfonts}
\usepackage{authblk} 
\usepackage{sectsty}
\usepackage{amsmath,amssymb,epsfig}
\usepackage{amscd}
\usepackage{amsthm}
\usepackage{mathrsfs}
\usepackage{dsfont}
\usepackage[applemac]{inputenc}
\usepackage[english]{babel}
\usepackage{enumitem} 
\usepackage[]{latexsym}
\usepackage{caption}
\usepackage{subfigure}
\usepackage{blindtext}
\usepackage{xcolor}
\usepackage{verbatim}
\usepackage{cancel}

\newcommand*{\affaddr}[1]{#1}
\newcommand*{\affmark}[1][*]{\textsuperscript{#1}}

\newtheorem*{proof*}{Proof}

\newcommand{\be}{\begin{equation}}

\newcommand{\ee}{\end{equation}}
\def\beqa{\begin{eqnarray}}
\def\eeqa{\end{eqnarray}}
\def\bean{\begin{eqnarray*}}
\def\eean{\end{eqnarray*}}

\renewenvironment{thebibliography}[1]
         {\section*{References}\frenchspacing\small
          \begin{list}{[\arabic{enumi}]}
         {\usecounter{enumi}\parsep=2pt\topsep 0pt
         \settowidth{\labelwidth}{[#1]}
         \leftmargin=\labelwidth\advance\leftmargin\labelsep
         \rightmargin=0pt\itemsep=1pt\sloppy}}{\end{list}}

 \numberwithin{equation}{section}

\input xy
\xyoption{all}

\usepackage[normalem]{ulem}

\newcommand{\ket}[1]{\left| #1 \right\rangle}

\newcommand{\braket}[2]{\left\langle \vphantom {#1 #2} #1 \hphantom{|} \right| \left. \vphantom {#1 #2} #2 \right\rangle}

\newcommand{\braopket}[3]{\left\langle \vphantom {#1 #2 #3} #1 \hphantom{|} \right| #2 \left| \hphantom{|} \vphantom {#1 #2 #3} #3 \right\rangle}

\newcommand{\ba}{\begin{eqnarray}}
\newcommand{\ea}{\end{eqnarray}}

\title{\textbf{\textsf{A note on the scalar products in sLQC}}\vspace{0.35cm}}

\author{
\textsf{Norbert Bodendorfer\affmark[1]\footnote{\texttt{norbert.bodendorfer@physik.uni-r.de}}}\\
\affaddr{\affmark[1]\textsf{Institute for Theoretical Physics, University of Regensburg,}}\\
\affaddr{\textsf{93040 Regensburg, Germany}}\vspace{-0.5cm}
}

\begin{document}

\maketitle

\begin{abstract}
\textsf{Solvable loop quantum cosmology provides a simple model of spatially flat, homogeneous, and isotropic quantum cosmology where the initial singularity is resolved. A close inspection of the literature reveals that there exist two different proposals for a scalar product that are usually assumed to be identical, but agree only in the large volume limit. The small volume limit, and thus the reported difference, turns out to be important for questions such as coarse graining fundamental quantum states in full theory embeddings of these cosmological models. }
\end{abstract}

\section{Introduction}

Solvable loop quantum cosmology \cite{AshtekarRobustnessOfKey} is a model for spatially flat, homogeneous, and isotropic quantum cosmology that can be solved analytically due to a specific choice of factor ordering and lapse. It reproduces earlier numerical findings \cite{AshtekarQuantumNatureOfAnalytical} to great accuracy for large spatial volumes and differs w.r.t. the model in \cite{AshtekarQuantumNatureOfAnalytical} mainly by neglecting inverse triad corrections that become relevant for small volumes (as compared to the Planck scale modulo a possible rescaling by the Barbero-Immirzi parameter). Since the spatial volume is naturally taken to be large in cosmological applications, these differences are usually not relevant and an effective classical description is very accurate in this regime \cite{AshtekarQuantumNatureOfAnalytical}. 

Recently, a strategy to embed this model into a full theory context was devised \cite{BVI}, based on choosing full theory variables that are maximally close to those used in sLQC. A full theory quantum state is approximated as a product state of $N$ copies (cells) of sLQC states, which due to the homogeneity and isotropy assumptions are chosen to be identical. This construction allows to ask the question of coarse graining or refining such quantum states which turns out to be identical to the issue of fiducial cell independence in LQC \cite{BodendorferStateRefinementsAnd}. After a sufficient amount of refinement moves, i.e. analogues of inverse block spin transformations, the spatial volume in a single cell can become arbitrarily small. As opposed to the above discussion, this puts corrections due to small volume, in particular so called inverse volume corrections, and their dynamical consequences into focus (see \cite{BojowaldTheBKLScenario} for a general discussion). In particular, it is important to ask whether these are irrelevant and die off after a sufficient amount of coarse graining (block-spin transformations), or whether they have consequences even at large volume. 

In this note, we will not deal with inverse volume corrections, but go a step back and point out that already at the level of scalar products there seem to be two different proposals in the literature. We will highlight where this difference comes from and how it disappears in the large volume limit where most of the previous work has taken place.

\section{sLQC from the volume representation} \label{sec:2}

The most straight forward way to derive sLQC is to contemplate about a consistent quantum dynamics of spatially flat, homogeneous, and isotropic quantum cosmology in the presence of a spatial volume that is quantised in integer multiples of the Planck volume\footnote{For simplicity, we will neglect the Barbero-Immirzi parameter which sets the discreteness scale w.r.t. the Planck scale in this paper, but all results also follow with it.}. Then, the trace of the extrinsic curvature, $b$, which is canonically conjugate to the signed spatial volume $v$, i.e $\{v,b \} = 1$ in units of $\hbar = c =12 \pi G =1$, cannot exist as an operator, whereas 
\be
	\widehat {e^{- i n b}} \ket{v} = \ket{v + n}, ~~ ~~n \in \mathbb Z
\ee   
is well-defined. Here, $\ket{v}$ is a volume eigenstate, i.e. $\hat v \ket{v} = v \ket{v}$. We note that $b$ is always implied to appear in exponentials as a suitable dimensionless ratio involving the Planck scale, which is set to one by our choice of natural units.

The scalar product between two states $ \ket{\tilde \chi_i} = \sum_{v \in \mathbb Z} \tilde \chi_i(v) \ket{v}$, $i = 1,2$, that implements the correct adjointness relations reads \cite{AshtekarRobustnessOfKey, AshtekarMathematicalStructureOf}
\be
	\braket{\tilde \chi_1}{\tilde \chi_2} = \sum_{v \in \mathbb Z} \overline{\tilde \chi_1(v)} \tilde \chi_2(v) \text{.} \label{eq:ScalarProductV}
\ee

The task is now to find a suitable factor ordering for the Hamiltonian constraint including matter degrees of freedom which leads to a tractable equation. For simplicity, one usually works with a massless scalar field $\phi$, where the classical Hamiltonian constraint reads $\mathcal H = \frac{p_\phi^2}{2v} - \frac{b^2 v}{2} \approx 0$. It is most straight-forward to consider $\phi$ as a physical time variable already classically and to quantise 
\be
	p_\phi = H_{\text{true}} := \sqrt{b^2 v^2}, 
\ee
where $H_{\text{true}}$ is the generator of $\phi$-time translations that acts only on the gravitational variables. This Hamiltonian can also be obtained by choosing the lapse $N = v$ and quantising the full Hamiltonian constraint as in \cite{AshtekarRobustnessOfKey}. 

The above observation forces us to substitute $b$ by a function of exponentials $e^{-i n b}$ in the process of quantising operators.
It turns out that the choice
\be
	\hat H_{\text{true}}^2 = \widehat{\sqrt{|v|}} \widehat{\sin( b)} \widehat{|v|} \widehat{\sin( b)} \widehat{\sqrt{|v|}} \label{eq:HtrueDef}
\ee
leads to a tractable Schr\"odinger equation and a) generates finite translations in $v$, b) annihilates the zero volume state, c) preserves the lattices with support on $v_n = c + 2  n $, $c \in \{0, 1\}$, $n \in \mathbb Z$, d) preserves the subsets of states with only positive or negative volumes, e) commutes with $v \mapsto -v$, and f) features departures from the classical theory only for matter energy densities $p_\phi^2 / (2v^2)$ close to the Planck density. In a quantisation involving also $\phi$, these properties can be considered as a consequence of the lapse choice $N = v$ along with the above factor ordering.

We restrict to $c = 0$. 
Due to the decoupling of the zero volume state, we can rescale our wave functions as 
\be
	 \tilde \chi(v) = \sqrt{|v|} \tilde \psi(v) ~~ \text{for}  ~~ v \neq 0 ~~
\ee
leading to 
\be
	\hat H_{\text{true}}^2 \tilde \psi(v) =    \widehat{\sin( b)} \widehat{|v|} \widehat{\sin( b)} \widehat{|v|} \, \tilde \psi(v) \label{eq:HamResc}
\ee
and 
\be
	\braket{\tilde \psi_1}{\tilde \psi_2} = \sum_{v \in \mathbb Z} \overline{\tilde \psi_1(v)} |v| \tilde \psi_2(v) \text{.} \label{eq:ScalarProductChiV}
\ee
While $\hat H_{\text{true}}^2$ as expressed in \eqref{eq:HamResc} now only preserves semi-positive / negative volumes, the zero-volume states in $\tilde \psi(v)$ produced by the action of \eqref{eq:HamResc} are annihilated once transforming back to $\tilde \chi(v)$. At the same time, the zero volume state in $\tilde \psi(v)$ is still annihilated by \eqref{eq:HamResc}, and thus does not dynamically influence non-zero volume states and decouples as before. Accordingly, we can drop the absolute values around both $v$ in the operator.

We perform a Fourier transform on the lattice $2 \mathbb Z$, our chosen dynamical subsector, 
\be
	 \psi(b) = \sum_{v \in 2  \mathbb Z} \tilde \psi(v) e^{-i b v}, ~~  ~~~~ \tilde \psi(v) = \frac{ 1}{\pi} \int_{0}^{\pi} db \, e^{i  v b} {\psi}(b) 
\ee
which results in 
\be
	\hat H_{\text{true}}^2 \, \psi(b) = -  \sin( b) \partial_b \sin( b) \partial_b \, \psi(b)
\ee
and the scalar product 
\be
	\braket{\psi}{\psi'}= \frac{1}{\pi} \int_0^{\pi} db \,\overline{ \psi(b)} |i \partial_b|  \psi'(b) \text{.} \label{eq:SPb}
\ee
To simplify the gravitational Hamiltonian further, we map the interval $(0, \pi)$ to $(-\infty, \infty)$ via 
\be
	x = \log \left( \tan ( b/2) \right)  ~~~ \Leftrightarrow ~~~ b = 2 \tan^{-1}(e^x) 
\ee
and compute
\be
	\partial_x =  \sin( b) \partial_b, ~~~~\partial_b =  \cosh(x) \partial_x, ~~~~ dx = \frac{1}{\sin( b)} db, ~~~~ db = \frac{1}{ \cosh x} dx \text{.}
\ee
Consequently, the gravitational Hamiltonian becomes 
\be
	\hat H_{\text{true}}^2 \, \psi(x) = -  \partial_x^2 \, \psi(x)
\ee
The boundary conditions $\psi(x=-\infty) = \psi(x=\infty)$ follow from $\psi(b=0) = \psi(b=\pi)$. In particular, they are satisfied for arbitrary superpositions of volume eigenstates $e^{-2 i v \text{ArcTan}(e^x) }$. 

The resulting Schr\"odinger equation describing evolution of quantum states in scalar field time reads
\be
	- i \partial_\phi \, \psi(x, \phi) = \sqrt{-\partial_x^2} \, \psi(x, \phi) \text{.}  \label{eq:HamQuantSimplPos}
\ee
Taking the positive square root of $-\partial_x^2$ corresponds to taking only positive frequency solutions. Such a choice ensures that the scalar product of this section, see e.g. \eqref{eq:ScalarProductV} or \eqref{eq:ScalarProductX}, is preserved in scalar field time (as explained after equation \eqref{eq:ScalarProductForm}). 

After another Fourier transform, we obtain the solutions
\begin{align}
	\psi_\text{phys}(x, \phi) &=   \int_{- \infty}^{ \infty} dk \, \tilde \psi_\text{phys}(k) e^{-ikx + i |k|(\phi - \phi_0)} \nonumber \\
			&=   \int_{- \infty}^{ 0} dk \,  \tilde \psi_\text{phys}(k) e^{-ik( x+\phi)} e^{  i k  \phi_0} +    \int_0^{ \infty} dk \, \tilde \psi_\text{phys}(k) e^{-ik(x-\phi )} e^{ - i k  \phi_0} \nonumber \\
			&=: \psi_L(x_+) + \psi_R(x_-) \text{.} \label{eq:ACSLimit}
\end{align}
For future reference, we denote the eigenfunctions of $i \partial_x$ with eigenvalue $k$ by
\ba
	 \psi_k(x) &=&e^{-ikx } \\
	   \psi_k(b) &=& e^{-i k \log(\tan( b/2))} \\
	 \tilde  \psi_k(v) &=&  \frac{ 1}{\pi} \int_{0}^{\pi} db \, e^{i  v b} e^{-i k \log(\tan( b/2))} \label{eq:ChiKOfV}
\ea
The scalar product in the $x$-representation reads
\ba
	\braket{\psi}{\psi'}&=&\frac{1}{\pi} \int_{-\infty}^{\infty} dx \frac{1}{\cosh(x)} \,\overline{ \psi(x)} |i \cosh(x)\partial_x|  \psi'(x)  \text{.} \label{eq:ScalarProductX}
\ea
The absolute value can be dealt with by decomposing $\psi_k(x)$ into two functions with support only on positive / negative volumes. The corresponding transformation was worked out in \cite{AshtekarCastingLoopQuantum} and reads ($k>0$)
\ba
	\psi^+_k &=& \frac{1}{2} \left(e^{\pi k/2} \psi_k + e^{-\pi k/2} \psi_{-k} \right) \label{eq:PsiPDef}\\
	\psi^-_k &=& \frac{1}{2} \left(e^{-\pi k/2} \psi_k + e^{\pi k/2} \psi_{-k} \right) \label{eq:PsiMDef}
\ea
with inverse transform
\ba	
	\psi_k &=& \frac{1}{\sinh(\pi k)} \left( e^{\pi k /2} \psi^+_k-e^{-\pi k /2} \psi^-_k \right) \label{eq:LQCBackTransform1}\\
	\psi_{-k} &=& \frac{1}{\sinh(\pi k)} \left(-e^{-\pi k /2} \psi^+_k+e^{\pi k /2} \psi^-_k \right) \label{eq:LQCBackTransform2} \text{.}
\ea
This result is non-trivial and requires to combine the integrals in \eqref{eq:ChiKOfV} from both contributions to a contour integral. The resulting expressions in the volume basis are \cite{AshtekarCastingLoopQuantum, CraigDynamicalEigenfunctionsAnd}
\ba
	 \tilde \psi^+_k(2n) &=& I(k, n), ~~~ \tilde \psi^-_k(2n) = I(k, -n), ~~ \text{with} ~ n \in \mathbb Z ~ \text{and} \\
	\nonumber && \\
	 I(k,n) &=& 
		\begin{cases}
		    \frac{1}{(2n)!} \left. \left( \frac{d}{ds} \right)^{2n} \right|_{s=0} \left( \frac{1-s}{1+s} \right)^{-ik}& \text{if } n \geq 0\\
 		   0              & \text{if } n<0
		\end{cases}\\
		&=&
		\begin{cases}
		   - i k \frac{\Gamma(2n - i k)}{\Gamma(1+2n)\Gamma(1- ik)} {}_2F_1(ik, -2n; 1-2n+ik; 1) & \text{if } n \geq 0\\
 		   0              & \text{if } n<0
		\end{cases}
\ea
The scalar products can now be evaluated as
\ba
	\braket{\psi^\pm_k}{\psi^\pm_{k'}} &=&  k \sinh(\pi k) \,  \delta(k,k') \label{eq:SPS1}\\
	\braket{\psi^\pm_k}{\psi^\mp_{k'}} &=& 0 \label{eq:SPS2}
\ea
and by \eqref{eq:LQCBackTransform1}, \eqref{eq:LQCBackTransform2}
\ba
	\braket{\psi_{\pm k}}{\psi_{\pm k'}} &=& 2  k \coth(\pi k) \times \delta(k,k') \label{eq:LQCScalarProduct1}\\
	\braket{\psi_{\pm k}}{\psi_{\mp k'}}&=& - 2  \frac{k}{\sinh(\pi k)} \times \delta(k, k') \label{eq:LQCScalarProduct2}  \text{.}
\ea
Due to the form
\be
	\braket{\psi}{\psi'} = \int_{-\infty}^{\infty} dk \, \overline{\psi(k)} \left( g_1(|k|) \delta(k, k') + g_2(|k|) \delta(k, -k') \right) \psi'(k')   \label{eq:ScalarProductForm}
\ee 
of the scalar product, it is preserved under $\phi$-time evolution generated by $\sqrt{-\partial_x^2} $ as the $\phi$-dependence from \eqref{eq:ACSLimit} cancels. This follows from the modulus of $k$ in the positive frequency restriction $-i \partial_\phi = |k|$ (otherwise, the contribution from $ g_2(|k|) \delta(k, -k')$ would be $\phi$-dependent).  

Due to our deparametrisation point of view, $v$ and $\sin(b)$ are physical observables. Similarly, the usual LQC observables, e.g. $v(\tilde \phi)$ and $\sin(b(\tilde \phi))$ where $\tilde \phi$ is a chosen reference scalar field time, could be constructed if \eqref{eq:LQCScalarProduct1}, \eqref{eq:LQCScalarProduct2} were constructed via a group averaging procedure starting from a quantisation that also includes $\phi$, as e.g. in \cite{AshtekarRobustnessOfKey}. In this case, one would start with a scalar product that additionally integrates over $\phi$ and formally insert $\delta(\hat H)$, which removes the $\phi$ integration while setting $\omega = |k|$, where $\omega$ is the analog of $k$ in the Fourier transform of the $\phi$ sector.

\section{Alternative scalar product}

The above derivation was originally given in \cite{AshtekarRobustnessOfKey} up to \eqref{eq:ACSLimit}. The physical scalar product was then chosen as the standard Klein-Gordon scalar product 
\ba
	\braket{\psi}{\psi'}_{\text{ACS}}&=&2 \int_{-\infty}^{\infty} dk  \,\overline{ \tilde \psi(k)} |k|  \tilde \psi'(k)  \label{eq:ACSSPk} \\
							&=&\frac{1}{\pi} \int_{-\infty}^{\infty} dx  \,\overline{ \psi(x)} |i \partial_x|  \psi'(x)  \label{eq:ACSSPx}
\ea
and it was mentioned that it would follow from a group averaging procedure. The scalar product that the group averaging procedure should have started with, in particular its relation to \eqref{eq:ScalarProductV}, was not provided. We note that \eqref{eq:ACSSPk} can be obtained as the large $|k|$ limit of \eqref{eq:LQCScalarProduct1}, \eqref{eq:LQCScalarProduct2}, where $\coth(\pi k) \rightarrow 1$ and $\frac{k}{\sinh (\pi k)} \rightarrow 0$. This limit coincides with the large volume limit as can be most easily checked when considering coherent states, see e.g. \cite{CorichiCoherentSemiclassicalStates}.
The main qualitative difference is that left- and rightmovers are orthogonal w.r.t. \eqref{eq:ACSSPk}, while this holds only asymptotically for large $|k|$ for \eqref{eq:LQCScalarProduct2}.  

One may ask whether the observed differences vanish once one restricts to volume-symmetric states as usually done in sLQC, but the answer turns out to be negative. For this, it is sufficient to restrict to the positive volume sector and compute
\ba
	 \braket{\psi^+_k}{\psi{}^+_{k'}}_{\text{ACS}}&=& |k| \cosh(\pi k) \delta(k, k') 
\ea
using the definitions \eqref{eq:PsiPDef}, \eqref{eq:PsiMDef}. One sees again that the scalar product \eqref{eq:SPS1} from section \ref{sec:2} is reproduced only for large $k$. Similarly, $\braket{\psi^+_k}{\psi^-_k}_{\text{ACS}} \neq 0$ unless one takes again the large $k$ limit and normalizes the states. 

A related difference occurs in the volume. 
Since the candidate volume operator $i \partial_b = i \cosh(x) \partial_x$ that is suggested by the derivation of \eqref{eq:HamQuantSimplPos} (assuming that one had \eqref{eq:ScalarProductV} in mind as a starting scalar product) is not self-adjoint w.r.t. \eqref{eq:ACSSPx}, its self-adjoint part w.r.t. \eqref{eq:ACSSPx} was used in \cite{AshtekarRobustnessOfKey}\footnote{This procedure may be questioned on the ground that it is arbitrary to choose the new physical scalar product w.r.t. which one projects an operator that was constructed w.r.t. another scalar product. Rather, one could obtain the physical scalar product w.r.t. which the operators would already be self-adjoint via group averaging as in section \ref{sec:2}. Also, projecting on self-adjoint parts (which becomes necessary as \eqref{eq:ACSSPk} does not implement the correct acjointness relations for $v$ and $\sin(b)$) does generally not commute with taking commutators, i.e. it may change the commutation relations and therefore spoil the quantum theory.}.
This leads to a volume operator and its absolute value that preserve left- ($k<0$) and rightmovers ($k>0$), a key difference to the one from section \ref{sec:2}. 
No such property can be inferred from the absolute value of the volume operator in section \ref{sec:2}, which was self-adjoint w.r.t. \eqref{eq:ScalarProductV} from the start. In fact, 
$\tilde \psi_k(v)$ has support on both positive and negative volumes with the same functional dependence through $I(k,n)$, but different relative weights. Again, in the limit of large $|k|$, this difference disappears as the ratio of positive and negative volumes scales as $e^{\pi k}$.
For the matrix elements, one obtains (formally, since matrix elements of the volume involving single Fourier modes generally diverge)
\ba
	\braopket{\psi_k}{|\hat v|}{\psi_{-k'}} &=& - \braopket{\psi_k}{|\hat v|}{\psi_{k'}} \frac{\cosh(\pi (k-k')/2)}{\cosh(\pi (k+k')/2)}, ~~~~~ k, k' > 0 \nonumber \\ & \stackrel{k, k' \gg 1}{=} & - \braopket{\psi_k}{|\hat v|}{\psi_{k'}} e^{-\pi \, \text{max}(k, k')}.
\ea
We can understand the asymptotic agreement between both scalar products also in the $x$-representation. Since $\tilde \psi_k$ is peaked on positive / negative volumes for large $|k|$, one can drop the absolute value in \eqref{eq:ScalarProductX} for the sign of $k = i \partial_x$ so that both factors of $\cosh(x)$ cancel and \eqref{eq:ACSSPx} is obtained. 

Transforming back to $b,v$-variables, we obtain 
\ba
	\braket{\psi}{\psi'}_{\text{ACS}}&=&\frac{1}{\pi} \int_0^{\pi} \frac{db}{\sin b} \,\overline{ \psi(b)} |i \sin(b) \partial_b|  \psi'(b) \label{eq:BtoB} \\
							&=& \sum_{v, v' \in 2 \mathbb Z} \overline{\tilde \psi(v)} \tilde \psi'(v') \underbrace{\frac{1}{\pi} \int_0^\pi \frac{db}{\sin b} e^{i b v} |i \sin (b) \partial_b| e^{-i b v'}}_{K(v,v')}. \label{eq:BtoV}
\ea
The absolute value we encountered in \eqref{eq:ScalarProductX} in the x-representation now appears in the $b$-representation, leading to a non-trivial kernel $K(v,v')$ for the scalar product in the $v$-representation, as opposed to \eqref{eq:ScalarProductChiV}. In particular, $K(v,v')$ has support also for $v \neq v'$ as can be checked numerically. This last observation finishes the comparison of the two scalar products. 

We remark that one may construct an LQC model identical to the one from section \ref{sec:2} with scalar product \eqref{eq:ACSSPk} by transforming $\psi(b)$ and $\psi'(b)$ in \eqref{eq:BtoB} such that the new integration kernel reads $|i \partial_b |$ as in \eqref{eq:SPb}. Then, the rescaled wave functions have to be interpreted as those of \eqref{eq:SPb} and the operators $\hat v$ and $\widehat{\sin(b)}$ acting on them can be defined as in section \ref{sec:2}. They could then be expressed in the $k$ representation underlying \eqref{eq:ACSSPk}. We note two things: first, we didn't check whether this leads to the operators used in \cite{AshtekarRobustnessOfKey} in the $k$ representation. Second, even if it does, it changes the relation to the volume representation explained in section \ref{sec:2} due to the additional transformation that changes the integral kernel in \eqref{eq:BtoB} to $|i \partial_b|$.

\section{Conclusion}

We have pointed out in this note that the scalar product of sLQC originally proposed in \cite{AshtekarRobustnessOfKey} does not agree with the standard LQC one. This point seems to have remained unnoticed in the literature, whereas equality is usually assumed\footnote{For example, \cite{AshtekarCastingLoopQuantum} cites \cite{AshtekarRobustnessOfKey} for the scalar product \eqref{eq:SPb}, which is equivalent to \eqref{eq:ScalarProductV}, whereas \cite{AshtekarRobustnessOfKey} never provides this. In \cite{CraigDynamicalEigenfunctionsAnd} and \cite{CraigConsistentProbabilitiesInLoop}, one uses the scalar product \eqref{eq:ScalarProductV} while citing \cite{AshtekarRobustnessOfKey} for the model.}. Starting from the scalar products \eqref{eq:SPS1} and \eqref{eq:SPS2} which were originally derived in \cite{AshtekarCastingLoopQuantum}, the relevant computation to check this is just the simple inverse transform \eqref{eq:LQCBackTransform1}, \eqref{eq:LQCBackTransform2}, yielding \eqref{eq:LQCScalarProduct1}, \eqref{eq:LQCScalarProduct2}. To the best of our knowledge, these expressions (and their surrounding interpretation given in this paper) have not appeared in the literature before.  

From the point of view of a full theory embedding \cite{BVI}, the standard scalar product from section \ref{sec:2} is preferred as it directly follows from integrating point holonomies w.r.t. the corresponding Haar measure. In contrast, \eqref{eq:BtoV} does not appear to admit such an interpretation.

\section*{Acknowledgments}

The author was supported by an International Junior Research Group grant of the Elite Network of Bavaria. Several discussions with Parampreet Singh on the subject are gratefully acknowledged, as well as email exchanges with Miguel Campiglia and David Craig to confirm the novelty of the result.

%\bibliographystyle{utphysmendeley}
%\bibliography{library}

\end{document}